\documentclass[twocolumn,floatfix,eqsecnum,showpacs]{revtex4}
\usepackage{graphicx}
\usepackage{amsmath}
\usepackage{bm}
\usepackage{amssymb,amsfonts}
\usepackage{times}
\begin{document}

\title{Boundary conditions for Dirac fermions on a terminated honeycomb lattice}
\author{A. R. Akhmerov and C. W. J. Beenakker}
\affiliation{Instituut-Lorentz, Universiteit Leiden, P.O. Box 9506, 2300 RA Leiden, The Netherlands}
\date{October, 2007}
\begin{abstract}
We derive the boundary condition for the Dirac equation corresponding to a tight-binding model on a two-dimensional honeycomb lattice terminated along an arbitary direction. Zigzag boundary conditions result generically once the boundary is not parallel to the bonds. Since a honeycomb strip with zigzag edges is gapless, this implies that confinement by lattice termination does not in general produce an insulating nanoribbon. We consider the opening of a gap in a graphene nanoribbon by a staggered potential at the edge and derive the corresponding boundary condition for the Dirac equation. We analyze the edge states in a nanoribbon for arbitrary boundary conditions and identify a class of propagating edge states that complement the known localized edge states at a zigzag boundary.
\end{abstract}
\maketitle

\section{Introduction}
\label{intro}

The electronic properties of graphene can be described by a difference equation (representing a tight-binding model on a honeycomb lattice) or by a differential equation (the two-dimensional Dirac equation) \cite{Wal47,DiV84}. The two descriptions are equivalent at large length scales and low energies, provided the Dirac equation is supplemented by boundary conditions consistent with the tight-binding model. These boundary conditions depend on a variety of microscopic properties, determined by atomistic calculations \cite{Son06}.

For a general theoretical description, it is useful to know what boundary conditions on the Dirac equation are allowed by the basic physical principles of current conservation and (presence or absence of) time reversal symmetry --- independently of any specific microscopic input. This problem was solved in Refs.\ \cite{McC04,Akh07}. The general boundary condition depends on one mixing angle $\Lambda$ (which vanishes if the boundary does not break time reversal symmetry), one three-dimensional unit vector $\bm{n}$ perpendicular to the normal to the boundary, and one three-dimensional unit vector $\bm{\nu}$ on the Bloch sphere of valley isospins. Altogether, four real parameters fix the boundary condition.

In the present paper we investigate how the boundary condition depends on the crystallographic orientation of the boundary. As the orientation is incremented by $30^{\circ}$ the boundary configuration switches from armchair (parallel to one-third of the carbon-carbon bonds) to zigzag (perpendicular to another one-third of the bonds). The boundary conditions for the armchair and zigzag orientations are known \cite{Bre06}. Here we show that the boundary condition for intermediate orientations remains of the zigzag form, so that the armchair boundary condition is only reached for a discrete set of orientations. 

Since the zigzag boundary condition does not open up a gap in the excitation spectrum \cite{Bre06}, the implication of our result (not noticed in earlier studies \cite{Eza06}) is that a terminated honeycomb lattice of arbitrary orientation is metallic rather than insulating. We present tight-binding model calculations to show that, indeed, the gap $\Delta\propto\exp[-f(\varphi)W/a]$ in a nanoribbon at crystallographic orientation $\varphi$ vanishes exponentially when its width $W$ becomes large compared to the lattice constant $a$, characteristic of metallic behavior. The $\Delta\propto 1/W$ dependence characteristic of insulating behavior requires the special armchair orientation ($\varphi$ a multiple of $60^{\circ}$), at which the decay rate $f(\varphi)$ vanishes.

Confinement by a mass term in the Dirac equation does produce an excitation gap regardless of the orientation of the boundary. We show how the infinite-mass boundary condition of Ref.\ \cite{Ber87} can be approached starting from the zigzag boundary condition, by introducing a local potential difference on the two sublattices in the tight-binding model. Such a staggered potential follows from atomistic calculations \cite{Son06} and may well be the origin of the insulating behavior observed experimentally in graphene nanoribbons \cite{Che07,Han07}.

The outline of this paper is as follows. In Sec.\ \ref{generalbc} we formulate, following Refs.\ \cite{McC04,Akh07}, the general boundary condition of the Dirac equation on which our analysis is based. In Sec.\ \ref{latticeterm} we derive from the tight-binding model the boundary condition corresponding to an arbitrary direction of lattice termination. In Sec.\ \ref{staggered} we analyze the effect of a staggered boundary potential on the boundary condition. In Sec.\ \ref{spectrum} we calculate the dispersion relation for a graphene nanoribbon with arbitrary boundary conditions. We identify dispersive (= propagating) edge states which generalize the known dispersionless (= localized) edge states at a zigzag boundary \cite{Nak96}. The exponential dependence of the gap $\Delta$ on the nanoribbon width is calculated in Sec.\ \ref{bandgap} both analytically and numerically. We conclude in Sec.\ \ref{conclusion}.

\section{General boundary condition}
\label{generalbc}

The long-wavelength and low-energy electronic excitations in graphene are described by the Dirac equation
\begin{equation}
H\Psi=\varepsilon\Psi \label{DirEq}
\end{equation}
with Hamiltonian
\begin{equation}
H=v\tau_0\otimes(\bm{\sigma}\cdot \bm{p}) \label{DirHam}
\end{equation}
acting on a four-component spinor wave function $\Psi$. Here $v$ is the Fermi velocity and $\bm{p}=-i\hbar\nabla$ is the momentum operator. Matrices $\tau_i, \sigma_i$ are Pauli matrices in valley space and sublattice space, respectively (with unit matrices $\tau_0,\sigma_0$). The current operator in the direction $\bm{n}$ is $\bm{n}\cdot \bm{J}=v\tau_0\otimes(\bm{\sigma} \cdot \bm{n})$.

The Hamiltonian $H$ is written in the valley isotropic representation of Ref.\ \cite{Akh07}. The alternative representation $H'=v\tau_{z}\otimes(\bm{\sigma}\cdot \bm{p})$ of Ref.\ \cite{McC04} is obtained by the unitary transformation
\begin{equation}
H'=UHU^{\dagger},\;\; U=\tfrac{1}{2}(\tau_{0}+\tau_{z})\otimes\sigma_{0}+\tfrac{1}{2}(\tau_{0}-\tau_{z})\otimes\sigma_{z}. \label{Udef}
\end{equation}

As described in Ref.\ \cite{McC04}, the general energy-independent boundary condition has the form of a local linear restriction on the components of the spinor wave function at the boundary:
\begin{equation}
\Psi = M \Psi.
\label{PsiMPsi}
\end{equation}
The $4\times 4$ matrix $M$ has eigenvalue $1$ in a two-dimensional subspace containing $\Psi$, and without loss of generality we may assume that $M$ has eigenvalue $-1$ in the orthogonal two-dimensional subspace. This means that $M$ may be chosen as a Hermitian and unitary matrix,
\begin{equation}
M=M^{\dagger},\; M^2=1. \label{ConditionM1}
\end{equation}

The requirement of absence of current normal to the boundary, 
\begin{equation}
\left\langle\Psi|\bm{n}_B \cdot \bm{J}|\Psi\right\rangle=0, \label{CurrentAbsence}
\end{equation}
with $\bm{n}_B$ a unit vector normal to the boundary and pointing outwards, is equivalent to the requirement of anticommutation of the matrix $M$ with the current operator, 
\begin{equation}
\left\{M,\bm{n}_B \cdot \bm{J}\right\}=0. \label{ConditionMJ}
\end{equation}
That Eq.\ \eqref{ConditionMJ} implies Eq.\ \eqref{CurrentAbsence} follows from $\langle\Psi|\bm{n}_{B}\cdot\bm{J}|\Psi\rangle=\langle\Psi|M(\bm{n}_{B}\cdot\bm{J})M|\Psi\rangle=-\langle\Psi|\bm{n}_{B}\cdot\bm{J}|\Psi\rangle$. The converse is proven in App.\ \ref{CondCalc}.

We are now faced with the problem of determining the most general $4\times 4$ matrix $M$ that satisfies Eqs.\ \eqref{ConditionM1} and \eqref{ConditionMJ}. Ref.\ \cite{McC04} obtained two families of two-parameter solutions and two more families of three-parameter solutions. These solutions are subsets of the single four-parameter family of solutions obtained in Ref.\ \cite{Akh07},
\begin{equation}
M = \sin\Lambda\;\tau_0 \otimes (\bm{n}_1\cdot \bm{\sigma}) + \cos\Lambda\;(\bm{\nu} \cdot \bm{\tau})\otimes (\bm{n}_2\cdot \bm{\sigma}) \label{MainCond},
\end{equation}
where $\bm{\nu},\bm{n}_1,\bm{n}_2$ are three-dimensional unit vectors, such that $\bm{n}_1$ and $\bm{n}_2$ are mutually orthogonal and also orthogonal to $\bm{n}_B$. A proof that \eqref{MainCond} is indeed the most general solution is given in App.\ \ref{CondCalc}. One can also check that the solutions of Ref.\ \cite{McC04} are subsets of $M'=UMU^{\dagger}$. 

In this work we will restrict ourselves to boundary conditions that do not break time reversal symmetry. The time reversal operator in the valley isotropic representation is
\begin{equation}
T=-(\tau_y\otimes\sigma_y){\cal C},\label{TimeRev}
\end{equation}
with ${\cal C}$ the operator of complex conjugation. The boundary condition preserves time reversal symmetry if $M$ commutes with $T$. This implies that the mixing angle $\Lambda=0$, so that $M$ is restricted to a three-parameter family,
\begin{equation}
M=(\bm{\nu}\cdot\bm{\tau})\otimes(\bm{n}\cdot\bm{\sigma}),\;\; \bm{n}\perp \bm{n}_B.\label{CondTime}
\end{equation}

\section{Lattice termination boundary}
\label{latticeterm}
The honeycomb lattice of a carbon monolayer is a triangular lattice (lattice constant $a$) with two atoms per unit cell, referred to as $A$ and $B$ atoms (see Fig.\ \ref{Fig:Boundary}a). The $A$ and $B$ atoms separately form two triangular sublattices. The $A$ atoms are connected only to $B$ atoms, and vice versa. The tight-binding equations on the honeycomb lattice are given by
\begin{equation}
\begin{split}
\varepsilon\psi_A(\bm r)&=t[\psi_B(\bm r)+\psi_B(\bm r -\bm R_1)+\psi_B(\bm r -\bm R_2)],\\
\varepsilon\psi_B(\bm r)&=t[\psi_A(\bm r)+\psi_A(\bm r +\bm R_1)+\psi_A(\bm r +\bm R_2)].
\end{split}
\label{eq:TBHam}
\end{equation}
Here $t$ is the hopping energy, $\psi_A(\bm r)$ and $\psi_B(\bm r)$ are the electron wave functions on $A$ and $B$ atoms belonging to the same unit cell at a discrete coordinate $\bm r$, while $\bm R_1=(a\sqrt{3}/2,-a/2)$, $\bm R_2=(a\sqrt{3}/2,a/2)$ are lattice vectors as shown in Fig.\ \ref{Fig:Boundary}a. 

	Regardless of how the lattice is terminated, Eq.\ \eqref{eq:TBHam} has the electron-hole symmetry $\psi_B\to -\psi_B$, $\varepsilon \to -\varepsilon$.
For the long-wavelength Dirac Hamiltonian \eqref{DirHam} this symmetry is translated into the anticommutation relation 
\begin{equation}
H\sigma_z\otimes\tau_z+\sigma_z\otimes\tau_z H=0.	
	\label{eq:ehsym}
\end{equation}
 Electron-hole symmetry further restricts the boundary matrix $M$ in Eq.\ \eqref{CondTime} to two classes: zigzag-like ($\bm \nu=\pm \hat{\bm z}$, $\bm n=\hat{z}$) and armchair-like ($\bm \nu_z=\bm n_z=0$). In this section we will show that the zigzag-like boundary condition applies generically to an arbitrary orientation of the lattice termination. The armchair-like boundary condition is only reached for special orientations.

\subsection{Characterization of the boundary}
\begin{figure}[tbh]
\centerline{\includegraphics[width=\linewidth]{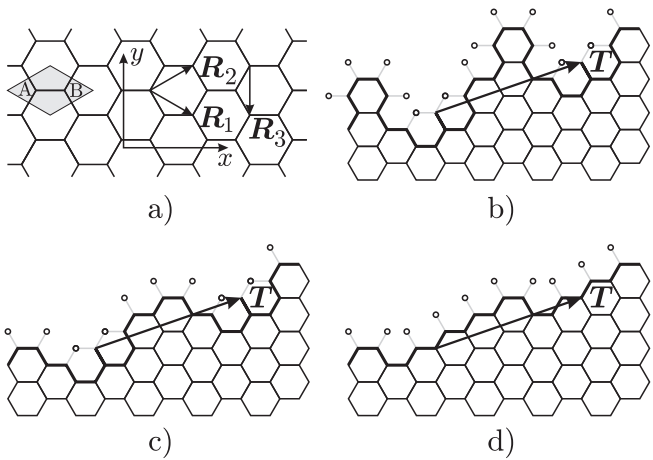}}
\caption{\label{Fig:Boundary}
(a) Honeycomb latice constructed from a unit cell (grey rhombus)
containing two atoms (labeled $A$ and $B$), translated over lattice
vectors $\bm{R}_{1}$ and $\bm{R}_{2}$. Panels b,c,d show three different
periodic boundaries with the same period
$\bm{T}=n\bm{R}_{1}+m\bm{R}_{2}$. Atoms on the boundary (connected by
thick solid lines) have dangling bonds (thin dotted line segments) to
empty neighboring sites (open circles). The number $N$ of missing sites
and $N'$ of dangling bonds per period is $\geq n+m$. Panel d shows a
{\em minimal\/} boundary, for which $N=N'=n+m$.
}
\end{figure}

A terminated honeycomb lattice consists of sites with three neighbors in the interior and sites with only one or two neighbors at the boundary. The absent neighboring sites are indicated by open circles in Fig.\ \ref{Fig:Boundary} and the dangling bonds by thin line segments. The tight-binding model demands that the wave function vanishes on the set of absent sites, so the first step in our analysis is the characterization of this set.
	We assume that the absent sites form a one-dimensional superlattice, consisting of a supercell of $N$ empty sites, translated over multiples of a superlattice vector $\bm T$. Since the boundary superlattice is part of the honeycomb lattice, we may write $\bm T=n \bm R_1 + m \bm R_2$ with $n$ and $m$ non-negative integers. For example, in Fig.\ \ref{Fig:Boundary} we have $n=1$, $m=4$. Without loss of generality, and for later convenience, we may assume that $m-n=0\ (\textrm{modulo }3)$. 
	
	The angle $\varphi$ between $\bm T$ and the armchair orientation (the $x$-axis in Fig.\ \ref{Fig:Boundary}) is given by 
\begin{equation}
	\varphi = \arctan \left(\frac{1}{\sqrt{3}}\frac{n-m}{n+m} \right),\ 
	-\frac{\pi}{6} \le \varphi\le\frac{\pi}{6}.
	\label{eq:alphadef}
\end{equation}
The armchair orientation corresponds to $\varphi=0$, while $\varphi=\pm\pi/6$ corresponds to the zigzag orientation. (Because of the $\pi/3$ periodicity we only need to consider $|\varphi|\le\pi/6$.)
	
	The number $N$ of empty sites per period $T$ can be arbitrarily large, but it cannot be smaller than $n+m$. Likewise, the number $N^{'}$ of dangling bonds per period cannot be smaller than $n+m$. We call the boundary \emph{minimal} if $N=N^{'}=n+m$. For example, the boundary in Fig.\ \ref{Fig:Boundary}d is minimal ($N=N^{'}=5$), while the boundaries in Figs.\ \ref{Fig:Boundary}b and \ref{Fig:Boundary}c are not minimal ($N=7, N^{'}=9$ and $N=5, N^{'}=7$, respectively). In what follows we will restrict our considerations to minimal boundaries, both for reasons of analytical simplicity \cite{note1} and for physical reasons (it is natural to expect that the minimal boundary is energetically most favorable for a given orientation).
	
	We conclude this subsection with a property of minimal boundaries that we will need later on. The $N$ empty sites per period can be divided into $N_A$ empty sites on sublattice $A$ and $N_B$ empty sites on sublattice $B$. A minimal boundary is constructed from $n$ translations over $\bm R_1$, each contributing one empty $A$ site, and $m$ translations over $\bm R_2$, each contributing one empty $B$ site. Hence, $N_A=n$ and $N_B=m$ for a minimal boundary.

\subsection{Boundary modes}

	The boundary breaks the two-dimensional translational invariance over $\bm R_1$ and $\bm R_2$, but a one-dimensional translational invariance over $T=n \bm R_1 + m \bm R_2$ remains. The quasimomentum $p$ along the boundary is therefore a good quantum number. The corresponding Bloch state satisfies
\begin{equation}
	\psi(\bm r+\bm T)= \exp(i k)\psi(\bm r),
	\label{eq:blochdef}
\end{equation} 
with $\hbar k = \bm {p}\cdot\bm{T}$.
	While the continuous quantum number $k \in (0, 2 \pi)$ describes the propagation along the boundary, a second (discrete) quantum number $\lambda$ describes how these boundary modes decay away from the boundary. We select $\lambda$ by demanding that the Bloch wave \eqref{eq:blochdef} is also a solution of 
\begin{equation}
\psi(\bm r+\bm R_3)=\lambda \psi(\bm r).	
	\label{eq:lambdadef}
\end{equation}
The lattice vector $\bm R_3=\bm R_1-\bm R_2$ has a nonzero component $a\cos\varphi>a \sqrt{3}/2$ perpendicular to $\bm T$. We need $|\lambda|\le 1$ to prevent $\psi(\bm r)$ from diverging in the interior of the lattice. The decay length $l_{\textrm{decay}}$ in the direction perpendicular to $\bm T$ is given by
\begin{equation}
l_{\textrm{decay}}=\frac{-a \cos\varphi}{\ln|\lambda|}.
	\label{eq:decaydef}
\end{equation}
	
	The boundary modes satisfying Eqs.\ \eqref{eq:blochdef} and \eqref{eq:lambdadef} are calculated in App.\ \ref{App:modes} from the tight-binding model. In the low-energy regime of interest (energies $\varepsilon$ small compared to $t$) there is an independent set of modes on each sublattice. On sublattice $A$ the quantum numbers $\lambda$ and $k$ are related by
\begin{subequations}
\label{eq:chareq}
\begin{equation}
	(-1-\lambda)^{m+n}=\exp(i k)\lambda^{n}
	\label{eq:chareqA}
\end{equation}
 and on sublattice B they are related by
\begin{equation}
	(-1-\lambda)^{m+n}=\exp(i k)\lambda^{m}.	
	\label{eq:chareqB}
\end{equation}
\end{subequations}		
	
	For a given $k$ there are ${\cal N}_A$ roots $\lambda_p$ of Eq. \eqref{eq:chareqA} having absolute value $\le 1$, with corresponding boundary modes $\psi_p$. We sort these modes according to their decay lengths from short to long, $l_{\textrm{decay}}(\lambda_p)\le l_{\textrm{decay}}(\lambda_{p+1})$, or $|\lambda_p|\le|\lambda_{p+1}|$. The wave function on sublattice $A$ is a superposition of these modes
\begin{equation}
	\psi^{(A)}=\sum_{p=1}^{{\cal N}_A} \alpha_p \psi_p,	
	\label{eq:psiA}
\end{equation}	
with coefficients $\alpha_p$ such that $\psi^{(A)}$ vanishes on the $N_A$ missing $A$ sites. Similarly there are ${\cal N}_B$ roots $\lambda^{'}_p$ of Eq.\ \eqref{eq:chareqB} with $|\lambda^{'}_p|\le 1$, $|\lambda^{'}_p|\le|\lambda^{'}_{p+1}|$. The corresponding boundary modes form the wave function on sublattice $B$,
\begin{equation}
	\psi^{(B)}=\sum_{p=1}^{{\cal N}_B} \alpha^{'}_p \psi^{'}_p,	
	\label{eq:psiB}
\end{equation}	with $\alpha^{'}_p$ such that $\psi^{(B)}$ vanishes on the $N_B$ missing $B$ sites.

\subsection{Derivation of the boundary condition}

	To derive the boundary condition for the Dirac equation it is sufficient to consider the boundary modes in the $k\to 0$ limit. The characteristic equations \eqref{eq:chareq} for $k=0$ each have a pair of solutions $\lambda_\pm = \exp(\pm 2 i \pi/3)$ that do not depend on $n$ and $m$. Since $|\lambda_\pm|=1$, these modes do not decay as one moves away from the boundary. The corresponding eigenstate $\exp(\pm i \bm K\cdot \bm r)$ is a plane wave with wave vector $\bm K=(4/3) \pi \bm R_3/a^2$. One readily checks that this Bloch state also satisfies Eq.\ \eqref{eq:blochdef} with $k=0$ [since $\bm K \cdot \bm T=2 \pi (n-m)/3=0\ (\textrm{modulo}\ 2 \pi)$].
	
	The wave functions \eqref{eq:psiA} and \eqref{eq:psiB} on sublattices $A$ and $B$ in the limit $k\to 0$ take the form
\begin{subequations}
\begin{align}
	\psi^{(A)}=\Psi_1 e^{i \bm K \cdot\bm r}+\Psi_4 e^{-i \bm K \cdot\bm r}+\sum_{p=1}^{{\cal N}_A-2} \alpha_p\psi_p,
\label{eq:psiA2}\\
	\psi^{(B)}=\Psi_2 e^{i \bm K \cdot\bm r}+\Psi_3 e^{-i \bm K \cdot\bm r}+\sum_{p=1}^{{\cal N}_B-2} \alpha_{p}^{'}\psi_p^{'}.
\label{eq:psiB2}
\end{align}
\label{eq:psi}
\end{subequations}
The four amplitudes ($\Psi_1$, $-i\Psi_2$, $i\Psi_3$, $-\Psi_4$) $\equiv \Psi$ form the four-component spinor $\Psi$ in the Dirac equation \eqref{DirEq}. The remaining ${\cal N}_A-2$ and ${\cal N}_B-2$ terms describe decaying boundary modes of the tight-binding model that are not included in the Dirac equation.
	
	We are now ready to determine what restriction on $\Psi$ is imposed by the boundary condition on $\psi^{(A)}$ and $\psi^{(B)}$. This restriction is the required boundary condition for the Dirac equation. In App.\ \ref{App:modes} we calculate that, for $k=0$,
\begin{align}
{\cal N}_A=	n - (n-m)/3+1,
	\label{eq:rootsA}\\
{\cal N}_B=	m - (m-n)/3+1,	
	\label{eq:rootsB}
\end{align}
so that ${\cal N}_A+{\cal N}_B=n+m+2$ is the total number of unknown amplitudes in Eqs.\ \eqref{eq:psiA} and \eqref{eq:psiB}. These have to be chosen such that $\psi^{(A)}$ and $\psi^{(B)}$ vanish on $N_A$ and $N_B$ lattice sites respectively. For the minimal boundary under consideration we have $N_A=n$ equations to determine ${\cal N}_A$ unknowns and $N_B=m$ equations to determine ${\cal N}_B$ unknowns.
	
	Three cases can be distinguished [in each case $n-m=0\ (\textrm{modulo } 3)$]:
\begin{enumerate}
	\item If $n>m$ then ${\cal N}_A \le n$ and ${\cal N}_B \ge m+2$, so $\Psi_1=\Psi_4=0$, while $\Psi_2$ and $\Psi_3$ are undetermined.
	\item If $n<m$ then ${\cal N}_B \le n$ and ${\cal N}_A \ge m+2$, so $\Psi_2=\Psi_3=0$, while $\Psi_1$ and $\Psi_4$ are undetermined.
	\item If $n=m$ then ${\cal N}_A = n+1$ and ${\cal N}_B = m+1$, so $|\Psi_1|=|\Psi_4|$ and $|\Psi_2|=|\Psi_3|$.
\end{enumerate}
In each case the boundary condition is of the canonical form $\Psi= (\bm\nu\cdot\bm\tau)\otimes(\bm n\cdot \bm\sigma) \Psi$ with
\begin{enumerate}
	\item $\bm\nu = -\hat{\bm z}$, $\bm n=\hat{\bm z}$ if $n>m$ (zigzag-type boundary condition).
	\item $\bm\nu = \hat{\bm z}$, $\bm n=\hat{\bm z}$ if $n<m$ (zigzag-type boundary condition).
	\item $\bm\nu\cdot\hat{\bm z} =0$, $\bm n\cdot\hat{\bm z}=0$ if $n=m$ (armchair-type boundary condition).
\end{enumerate}
	We conclude that the boundary condition is of zigzag-type for any orientation $\bm T$ of the boundary, unless $\bm T$ is parallel to the bonds [so that $n=m$ and $\varphi=0\ (\textrm{modulo }\pi/3)$].
\subsection{Precision of the boundary condition}

  At a perfect zigzag or armchair edge the four components of the Dirac spinor $\Psi$ are sufficient to meet the boundary condition. Near the boundaries with larger period and more complicated structure the wave function \eqref{eq:psi} also necessarily contains several boundary modes $\psi_p,\psi_p^{'}$ that decay away from the boundary. The decay length $\delta$ of the slowest decaying mode is the distance at which the boundary is indistinguishable from a perfect armchair or zigzag edge. At distances smaller than $\delta$ the boundary condition breaks down.

  In the case of an armchair-like boundary (with $n=m$), all the coefficients $\alpha_p$ and $\alpha_p^{'}$ in Eqs.\ \eqref{eq:psi} must be nonzero to satisfy the boundary condition. The maximal decay length $\delta$ is then equal to the decay length of the boundary mode $\psi_{n-1}$ which has the largest $|\lambda|$. It can be estimated from the characteristic equations \eqref{eq:chareq} that $\delta \approx |\bm T|$. Hence the larger the period of an armchair-like boundary, the larger the distance from the boundary at which the boundary condition breaks down.

  For the zigzag-like boundary the situation is different. On one sublattice there are more boundary modes than conditions imposed by the presence of the boundary and on the other sublattice there are less boundary modes than conditions. Let us assume that sublattice $A$ has more modes than conditions (which happens if $n<m$). The quickest decaying set of boundary modes sufficient to satisfy the tight-binding boundary condition contains $n$ modes $\psi_p$ with $p\le n$. The distance $\delta$ from the boundary within which the boundary condition breaks down is then equal to the decay length of the slowest decaying mode $\psi_n$ in this set and is given by
\begin{equation}
\delta=l_{\textrm{decay}}(\lambda_n)=-a \cos\varphi/\ln|\lambda_n|.
	\label{eq:delta}
\end{equation}
[See Eq.\ \eqref{eq:decaydef}.]

\begin{figure}
\includegraphics[width=0.8\linewidth]{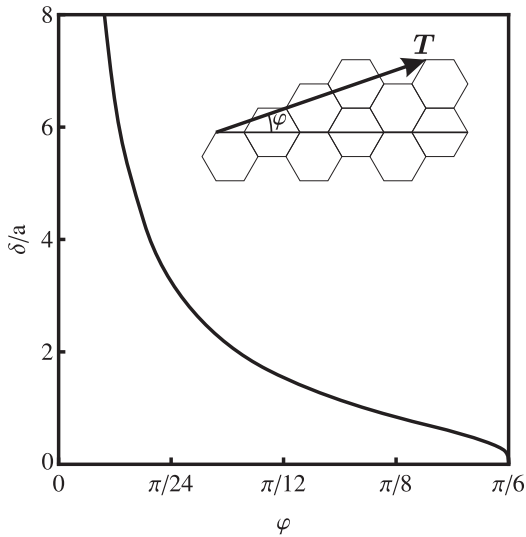}
\caption{\label{fig:l0} Dependence on the orientation $\varphi$ of the distance $\delta$ from the boundary within which the zigzag-type boundary condition breaks down. The curve is calculated from formula \eqref{eq:lambdan} valid in the limit $|\bm T|\gg a$ of large periods. The boundary condition becomes precise upon approaching the zigzag orientation $\varphi=\pi/6$.}
\end{figure}

  As derived in App.\ \ref{App:modes} for the case of large periods $|\bm T|\gg a$, the quantum number $\lambda_n$ satisfies the following system of equations:
\begin{subequations}
\begin{gather}
|1+\lambda_n|^{m+n}=|\lambda_n|^n,\\
\arg(1+\lambda_n)-\frac{n}{n+m} \arg(-\lambda_n)=\frac{n}{n+m} \pi.
\end{gather}
\label{eq:lambdan}
\end{subequations}
The solution $\lambda_n$ of this equation and hence the decay length $\delta$ do not depend on the length $|\bm T|$ of the period, but only on the ratio $n/(n+m)=(1-\sqrt{3}\tan\varphi)/2$, which is a function of the angle $\varphi$ between $\bm{T}$ and the armchair orientation [see Eq.\ \eqref{eq:alphadef}]. In the case $n>m$ when sublattice $B$ has more modes than conditions, the largest decay length $\delta$ follows upon interchanging $n$ and $m$.

As seen from Fig.\ \ref{fig:l0}, the resulting distance $\delta$ within which the zigzag-type boundary condition breaks down is zero for the zigzag orientation ($\varphi=\pi/6$) and tends to infinity as the orientation of the boundary approaches the armchair orientation ($\varphi=0$). (For finite periods the divergence is cut off at $\delta\sim |\bm T|\gg a$.) The increase of $\delta$ near the armchair orientation is rather slow: For $\varphi\gtrsim 0.1$ the zigzag-type boundary condition remains precise on the scale of a few unit cells away from the boundary.

Although the presented derivation is only valid for periodic boundaries and low energies, such that the wavelength is much larger than the length $|\bm{T}|$ of the boundary period, we argue that these conditions may be relaxed. Indeed, since the boundary condition is local, it cannot depend on the structure of the boundary far away, hence the periodicity of the boundary cannot influence the boundary condition. It can also not depend on the wavelength once the wavelength is larger than the typical size of a boundary feature (rather than the length of the period). Since for most boundaries both $\delta$ and the scale of the boundary roughness are of the order of several unit cells, we conclude that the zigzag boundary condition is in general a good approximation.

\subsection{Density of edge states near a zigzag-like boundary}

	A zigzag boundary is known to support a band of dispersionless states \cite{Nak96}, which are localized within several unit cells near the boundary. We calculate the 1D density of these edge states near an arbitrary zigzag-like boundary. Again assuming that the sublattice $A$ has more boundary modes than conditions ($n<m$), for each $k$ there are ${\cal N}_A(k)-N_A$ linearly independent states \eqref{eq:psiA}, satisfying the boundary condition. For $k\neq 0$ the number of boundary modes is equal to ${\cal N}_A = n - (m-n)/3$, so that for each $k$ there are 
\begin{equation}
N_{\textrm{states}}={\cal N}_A(k)-n=(m-n)/3
\end{equation}
edge states. The number of the edge states for the case when $n>m$ again follows upon interchanging $n$ and $m$. The density $\rho$ of edge states per unit length is given by
\begin{equation}
\label{eq:dos}
\rho=\frac{N_{\textrm{states}}}{|\bm T|}=\frac{|m-n|}{3 a \sqrt{n^2+nm+m^2}}=\frac{2}{3a}|\sin{\varphi}|.
\end{equation}
The density of edge states is maximal $\rho=1/3a$ for a perfect zigzag edge and it decreases continuously when the boundary orientation $\varphi$ approaches the armchair one. Eq. \eqref{eq:dos} explains the numerical data of Ref.\ \cite{Nak96}, providing an analytical formula for the density of edge states.

\section{Staggered boundary potential}
\label{staggered}
The electron-hole symmetry \eqref{eq:ehsym}, which restricts the boundary condition to being either of zigzag-type or of armchair-type, is broken by an electrostatic potential. Here we consider, motivated by Ref.\ \cite{Son06}, the effect of a staggered potential at the zigzag boundary. We show that the effect of this potential is to change the boundary condition in a continuous way from $\Psi=\pm\tau_z\otimes\sigma_z\Psi$ to $\Psi=\pm\tau_z\otimes (\bm\sigma\cdot[\hat{\bm z}\times \bm n_B])\Psi$. The first boundary condition is of zigzag-type, while the second boundary condition is produced by an infinitely large mass term at the boundary \cite{Ber87}.

\begin{figure}[tbh]
\centerline{\includegraphics[width=0.8\linewidth]{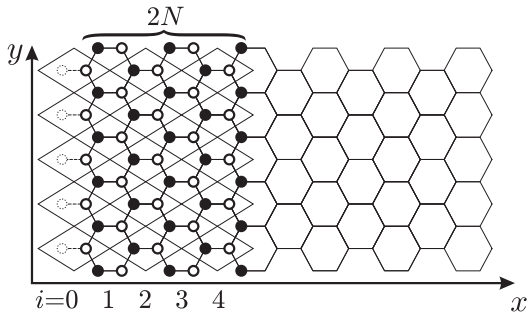}}
\caption{\label{stagpot}
Zigzag boundary with $V=+\mu$ on the $A$-sites (filled dots) and $V=-\mu$ on the $B$-sites (empty dots). The staggered potential extends over $2N$ rows of atoms nearest to the zigzag edge. The integer $i$ counts the number of unit cells away from the edge.
}
\end{figure}

The staggered potential consists of a potential $V_A=+\mu$, $V_B=-\mu$ on the $A$-sites and $B$-sites in a total of $2N$ rows closest to the zigzag edge parallel to the $y$-axis (see Fig.\ \ref{stagpot}). Since this potential does not mix the valleys, the boundary condition near a zigzag edge with staggered potential has the form 
\begin{equation}
\Psi=-\tau_z\otimes(\sigma_z\cos\theta+\sigma_y\sin\theta)\Psi,
	\label{eq:valleyBC}
\end{equation}
in accord with the general boundary condition \eqref{CondTime}. For $\theta=0,\pi$ we have the zigzag boundary condition and for $\theta=\pm\pi/2$ we have the infinite-mass boundary condition.
 
 To calculate the angle $\theta$ we substitute Eq.\ \eqref{eq:psi} into the tight-binding equation\ \eqref{eq:TBHam} (including the staggered potential at the left-hand side) and search for a solution in the limit $\varepsilon=0$. The boundary condition is precise for the zigzag orientation, so we may set $\alpha_{p}=\alpha'_{p}=0$. It is sufficient to consider a single valley, so we also set $\Psi_{3}=\Psi_{4}=0$. The remaining nonzero components are $\Psi_{1}e^{i\bm{K}\cdot\bm{r}}\equiv\psi_{A}(i)e^{iKy}$ and $\Psi_{2}e^{i\bm{K}\cdot\bm{r}}\equiv\psi_{B}(i)e^{iKy}$, where $i$ in the argument of $\psi_{A,B}$ numbers the unit cell away from the edge and we have used that $\bm{K}$ points in the $y$-direction. The resulting difference equations are
\begin{subequations}
\label{eq:TBstag}
\begin{gather}
-\mu \psi_A(i)=t[\psi_B(i)-\psi_B(i-1)],\ i=1,2,\ldots N,\\
\mu \psi_B(i)=t[\psi_A(i)-\psi_A(i+1)],\ i=0,1,2,\ldots N-1,\\
\psi_A(0)=0.
\end{gather}
\end{subequations}

\begin{figure}[tbh]
\includegraphics[width=0.8\linewidth]{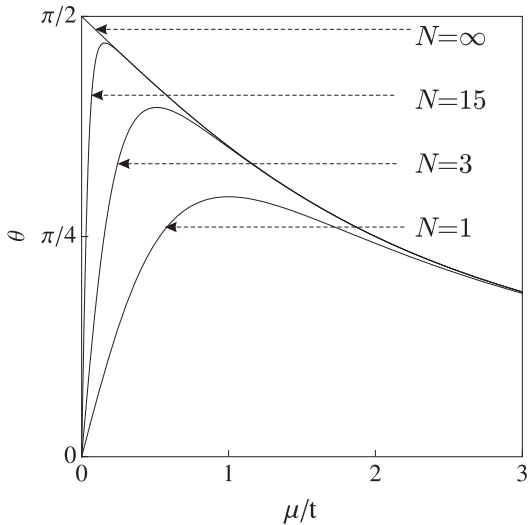}
\caption{\label{fig:inhom}
Plot of the parameter $\theta$ in the boundary condition \eqref{eq:valleyBC} at a zigzag edge with the staggered potential of Fig.\ \ref{stagpot}. The curves are calculated from Eq.\ \eqref{eq:resnz}. The values $\theta=0$ and $\theta=\pi/2$ correspond, respectively, to the zigzag and infinite-mass boundary conditions.
}
\end{figure}

For the $\Psi_{1},\Psi_{2}$ components of the Dirac spinor $\Psi$ the boundary condition \eqref{eq:valleyBC} is equivalent to
\begin{equation}
\psi_A(N)/\psi_B(N)=-\tan(\theta/2).
\label{eq:thetadef}
\end{equation}
Substituting the solution of Eq.\ \eqref{eq:TBstag} into Eq.\ \eqref{eq:thetadef} gives
\begin{equation}
\cos\theta=\frac{1+\sinh(\kappa)\sinh(\kappa+2 N \mu/t)}{\cosh(\kappa)\cosh(\kappa+2 N \mu/t)},
	\label{eq:resnz}
\end{equation}
with $\sinh\kappa=\mu/2t$. Eq.\ \eqref{eq:resnz} is exact for $N\gg 1$, but it is accurate within 2\% for any $N$. The dependence of the parameter $\theta$ of the boundary condition on the staggered potential strength $\mu$ is shown in Fig.\ \ref{fig:inhom} for various values of $N$. The boundary condition is closest to the infinite mass for $\mu/t\sim 1/N$, while the regimes $\mu/t\ll 1/N$ or $\mu/t \gg 1$ correspond to a zigzag boundary condition.

\section{Dispersion relation of a nanoribbon}
\label{spectrum}
A graphene nanoribbon is a carbon monolayer confined to a long and narrow strip. The energy spectrum $\varepsilon_{n}(k)$ of the $n$-th transverse mode is a function of the wave number $k$ along the strip. This dispersion relation is nonlinear because of the confinement, which also may open up a gap in the spectrum around zero energy. We calculate the dependence of the dispersion relation on the boundary conditions at the two edges $x=0$ and $x=W$ of the nanoribbon (taken along the $y$-axis).

	In this section we consider the most general boundary condition \eqref{CondTime}, constrained only by time-reversal symmetry. We do not require that the boundary is purely a termination of the lattice, but allow for arbitrary local electric fields and strained bonds. The conclusion of Sec.\ \ref{latticeterm}, that the boundary condition is either zigzag-like or armchair-like, does not apply therefore to the analysis given in this section.
	
	The general solution of the Dirac equation\ \eqref{DirEq} in the nanoribbon has the form $\Psi(x,y)=\Psi_{n,k}(x)e^{iky}$. We impose the general boundary condition \eqref{CondTime},
\begin{subequations}
\label{eq:bothBC}
\begin{gather}
\Psi(0,y)=(\bm \nu_1\cdot\bm\tau)\otimes(\bm n_1\cdot\bm\sigma) \Psi(0,y),\label{eq:LowerBC}\\
\Psi(W,y)=(\bm \nu_2\cdot\bm\tau)\otimes(\bm n_2\cdot\bm\sigma) \Psi(W,y),\label{eq:UpperBC}
\end{gather}
\end{subequations}
with three-dimensional unit vectors $\bm{\nu}_{i}$, $\bm{n}_{i}$, restricted by $\bm{n}_{i}\cdot\hat{\bm x}=0$ ($i=1,2$). (There is no restriction on the $\bm{\nu}_{i}$.)
Valley isotropy of the Dirac Hamiltonian \eqref{DirHam} implies that the spectrum does not depend on $\bm{\nu}_{1}$ and $\bm{\nu}_{2}$ separately but only on the angle $\gamma$ between them. The spectrum depends, therefore, on three parameters: The angle $\gamma$ and the angles $\theta_{1}$, $\theta_{2}$ between the $z$-axis and the vectors $\bm{n}_{1}$, $\bm{n}_{2}$. 

	The Dirac equation $H\Psi=\varepsilon\Psi$ has two plane wave solutions $\Psi\propto\exp(iky+iqx)$ for a given $\varepsilon$ and $k$, corresponding to the two (real or imaginary) transverse wave numbers $q$ that solve $(\hbar v)^{2}(k^{2}+q^{2})=\varepsilon^{2}$. Each of these two plane waves has a twofold valley degeneracy, so there are four independent solutions in total. Since the wavefunction in a ribbon is a linear combination of these four waves, and since each of the Eqs.\ (\ref{eq:LowerBC},\ref{eq:UpperBC}) has a two-dimensional kernel, these equations provide four linearly independent equations to determine four unknowns. The condition that Eq.\ \eqref{eq:bothBC} has nonzero solutions gives an implicit equation for the dispersion relation of the nanoribbon:
\begin{multline}
\cos\theta_1\cos\theta_2(\cos\omega-\cos^2\Omega)+\cos\omega\sin\theta_1\sin\theta_2\sin^2\Omega\\
-\sin\Omega[\sin\Omega\cos\gamma+\sin\omega\sin(\theta_1-\theta_2)]=0,
	\label{eq:RibDisp}
\end{multline}
where $\omega^2=4W^2[(\varepsilon/\hbar v)^2-k^2]$ and $\cos\Omega=\hbar v k/\varepsilon$. 

\begin{figure}[tbh]
\includegraphics[width=\linewidth]{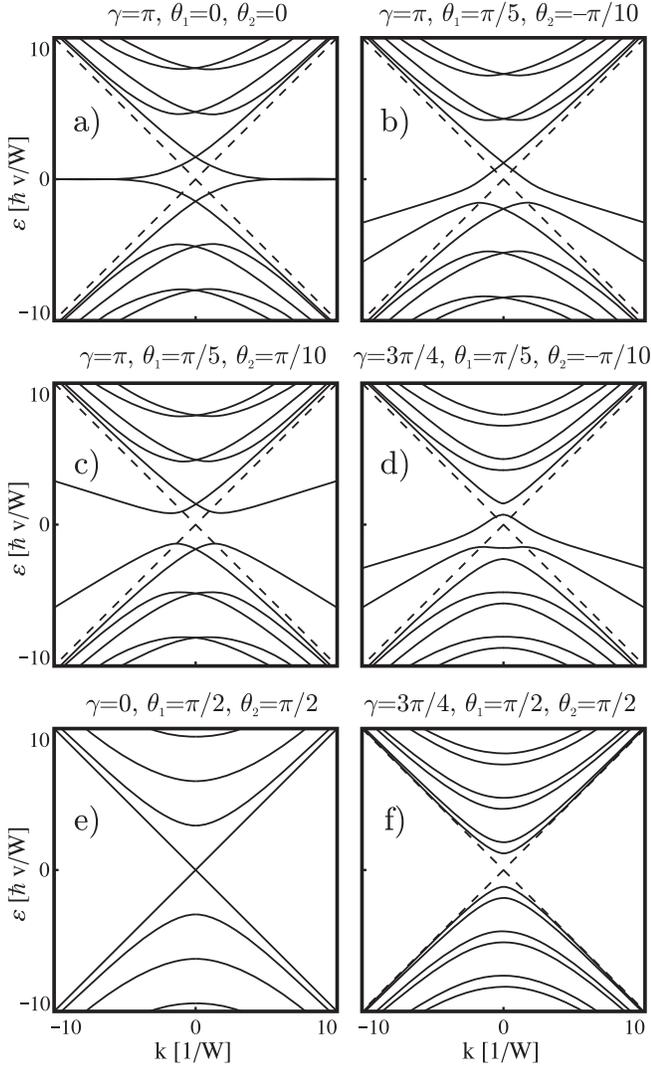}
	\caption{\label{fig:Dispersions}Dispersion relation of nanoribbons with different boundary conditions. The large-wave number asymptotes $|\varepsilon|=\hbar v|k|$ of bulk states are shown by dashed lines. Modes that do not approach these asymptotes are edge states with dispersion $|\varepsilon|=\hbar v|k \sin\theta_i|$. The zigzag ribbon with $\gamma=\pi$ and $\theta_1=\theta_2=0$ (a) exhibits dispersionless edge states at zero energy \cite{Nak96}. If $\theta_1$ or $\theta_2$ are nonzero (b, c) the edge states acquire linear dispersion and if $\sin\theta_1\sin\theta_2>0$ (c) a band gap opens. If $\gamma$ is unequal to $0$ or $\pi$ (d) the valleys are mixed which makes all the level crossings avoided and opens a band gap. Armchair-like ribbons with $\theta_1=\theta_2=\pi/2$ (e, f) are the only ribbons having no edge states.}
\end{figure}

	For $\theta_1=\theta_2=0$ and $\gamma=\pi$ Eq.\ (5.2) reproduces the transcendental equation of Ref.\ \onlinecite{Bre06} for the dispersion relation of a zigzag ribbon. In the case $\theta_{1}=\theta_{2}=\pi/2$ of an armchair-like nanoribbon, Eq.\ (5.2) simplifies to
\begin{equation}
	\cos\omega=\cos\gamma.
	\label{eq:ArmchairDispersion}
\end{equation}
This is the only case when the transverse wave function $\Psi_{n,k}(x)$ is independent of the longitudinal wave number $k$. In Fig.\ \ref{fig:Dispersions} we plot the dispersion relations for several different boundary conditions.

	The low energy modes of a nanoribbon with $|\varepsilon|<\hbar v |k|$ [see panels a-d of Fig.\ \ref{fig:Dispersions}] have imaginary transverse momentum since $q^2=(\varepsilon/\hbar v)^2-k^2<0$. If $|q|$ becomes larger than the ribbon width $W$, the corresponding wave function becomes localized at the edges of the nanoribbon and decays in the bulk. The dispersion relation \eqref{eq:RibDisp} for such an edge state simplifies to $\varepsilon= \hbar v |k| \sin\theta_1$ for the state localized near $x=0$ and $\varepsilon=-\hbar v |k| \sin\theta_2$ for the state localized near $x=W$. These dispersive edge states with velocity $v \sin\theta$ generalize the known \cite{Nak96} dispersionless edge states at a zigzag boundary (with $\sin\theta=0$).
	
	Inspection of the dispersion relation \eqref{eq:RibDisp} gives the following condition for the presence of a gap in the spectrum of the Dirac equation with arbitrary boundary condition: Either the valleys should be mixed ($\gamma\neq 0, \pi$) or the edge states at opposite boundaries should have energies of opposite sign ($\sin\theta_1\sin\theta_2>0$ for $\gamma=\pi$ or $\sin\theta_1\sin\theta_2<0$ for $\gamma=0$). 

\begin{figure}[tbh]
\includegraphics[width=0.8\linewidth]{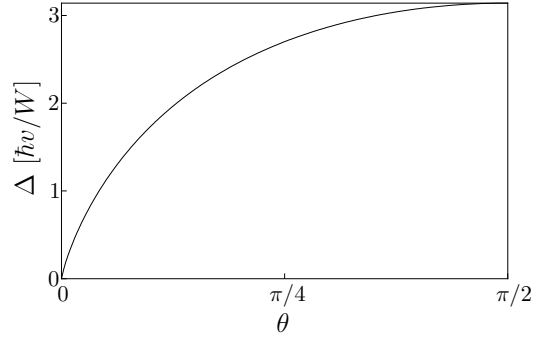}
	\caption{\label{fig:staggap}
Dependence of the band gap $\Delta$ on the parameter $\theta$ in the staggered potential boundary condition \eqref{eq:stagrib}.}
\end{figure}

 As an example, we calculate the band gap for the staggered potential boundary condition of Sec.\ \ref{staggered}. We assume that the opposite zigzag edges have the same staggered potential, so that the boundary condition is
\begin{subequations}\label{eq:stagrib}
\begin{gather}
\Psi(0,y)=+\tau_z\otimes(\sigma_z\cos\theta+\sigma_y\sin\theta) \Psi(0,y),\\
\Psi(W,y)=-\tau_z\otimes(\sigma_z\cos\theta+\sigma_y\sin\theta) \Psi(W,y).
\end{gather}
\end{subequations}
The dependence of $\theta$ on the parameters $\mu$, $N$ of the staggered potential is given by Eq.\ \eqref{eq:resnz}. This boundary condition corresponds to $\gamma=\pi$, $\theta_{1}=\theta_{2}=\theta$, so that it has a gap for any nonzero $\theta$. As shown in Fig.\ \ref{fig:staggap}, $\Delta(\theta)$ increases monotonically with $\theta$ from the zigzag limit $\Delta(0)=0$ to the infinite-mass limit $\Delta(\pi/2)=\pi\hbar v/W$.

\section{Band gap of a terminated honeycomb lattice}
\label{bandgap}
	In this section we return to the case of a boundary formed purely by termination of the lattice.
	A nanoribbon with zigzag boundary condition has zero band gap according to the Dirac equation (Fig.\ \ref{fig:Dispersions}a). According to the tight-binding equations there is a nonzero gap $\Delta$, which however vanishes exponentially with increasing width $W$ of the nanoribbon. We estimate the decay rate of $\Delta(W)$ as follows. 
	
	The low energy states in a zigzag-type nanoribbon are the hybridized zero energy edge states at the opposite boundaries. The energy $\varepsilon$ of such states may be estimated from the overlap between the edge states localized at the opposite edges, $\varepsilon=\pm (\hbar v/W)\exp(-W/l_\textrm{decay})$. In a perfect zigzag ribbon there are edge states with $l_\textrm{decay}=0$ (and $\varepsilon=0$), so that there is no band gap. For a ribbon with a more complicated edge shape the decay length of an edge state is limited by $\delta$, the length within which the boundary condition breaks down (see Sec. \ref{latticeterm}.D). This length scale provides the analytical estimate of the band gap in a zigzag-like ribbon:
	
\begin{equation}
	\Delta \sim \frac{\hbar v}{W} e^{-W /\delta},
	\label{eq:ZigzagGap}
\end{equation}
with $\delta$ given by Eqs.\ \eqref{eq:delta} and \eqref{eq:lambdan}.

	The band gap of an armchair-like ribbon is 
\begin{equation}
\Delta=(\hbar v/W)\arccos(\cos \gamma)	
	\label{eq:armdelta}
\end{equation}	
[see Eq.\ \eqref{eq:ArmchairDispersion} and panels e,f of Fig.\ \ref{fig:Dispersions}]. Adding another row of atoms increases the nanoribbon width by one half of a unit cell and increases $\gamma$ by $\bm K \cdot  \bm R_3=4\pi/3$, so the product $\Delta W$ in such a ribbon is an oscillatory function of $W$ with a period of 1.5 unit cells.

\begin{figure}[tbh]
\includegraphics[width=0.8\linewidth]{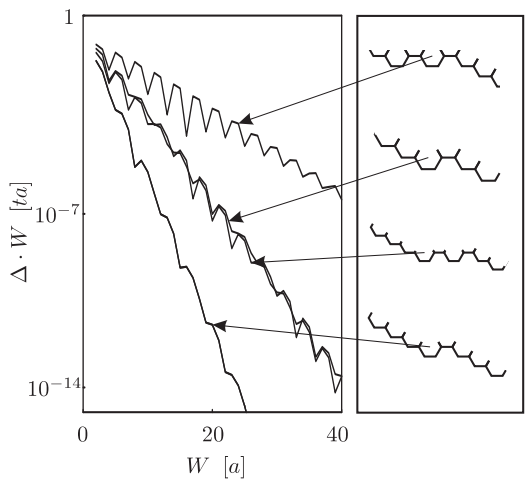}
\caption{\label{fig:Gap1}
Dependence of the band gap $\Delta$ of zigzag-like nanoribbons on the width $W$. The curves in the left panel are calculated numerically from the tight-binding equations. The right panel shows the structure of the boundary, repeated periodically along both edges.
}
\end{figure}

\begin{figure}[tbh]
\includegraphics[width=0.8\linewidth]{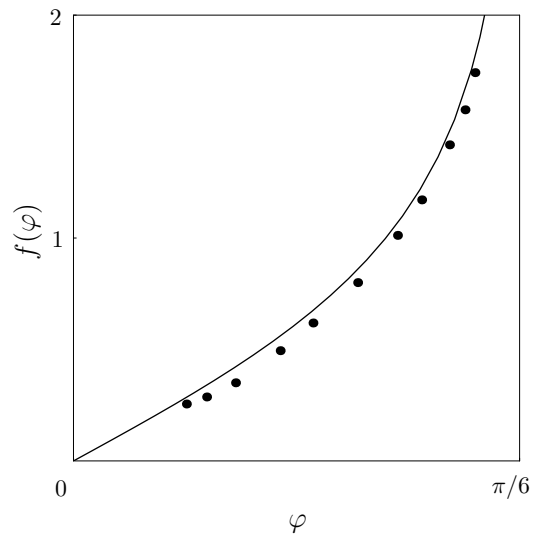}
\caption{\label{fig:Gap2}
Dependence of the gap decay rate on the orientation $\varphi$ of the boundary (defined in the inset of Fig.\ \ref{fig:l0}). The dots are the fits to numerical results of the tight-binding equations, the solid curve is the analytical estimate \eqref{eq:ZigzagGap}.
}
\end{figure}

	To test these analytical estimates, we have calculated $\Delta(W)$ numerically for various orientations and configurations of boundaries. As seen from Fig.\ \ref{fig:Gap1}, in ribbons with a non-armchair boundary the gap decays exponentially $\propto\exp[-f(\varphi)W/a]$ as a function of $W$. Nanoribbons with the same orientation $\varphi$ but different period $|\bm{T}|$ have the same decay rate $f$. As seen in Fig.\ \ref{fig:Gap2}, the decay rate obtained numerically agrees well with the analytical estimate $f=a/\delta$ following from Eq.\ \eqref{eq:ZigzagGap} (with $\delta$ given as a function of $\varphi$ in Fig.\ \ref{fig:l0}). The numerical results of Fig.\ \ref{fig:Gap1} are consistent with earlier studies of the orientation dependence of the band gap in nanoribbons \cite{Eza06}, but the exponential decrease of the gap for non-armchair ribbons was not noticed in those studies.
	
\begin{figure}[tbh]
\includegraphics[width=0.8\linewidth]{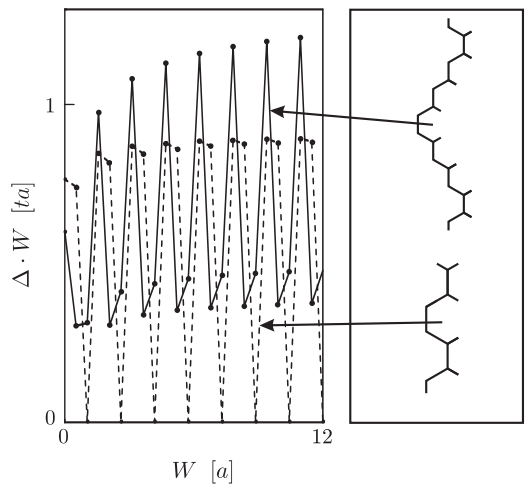}
\caption{\label{fig:armchair}
Dependence of the band gap $\Delta$ on the width $W$ for an armchair ribbon (dashed line)  and for a ribbon with a boundary of the same orientation but with a larger period (solid line). The curves are calculated numerically from the tight-binding equations.
}
\end{figure}

	For completeness we show in Fig.\ \ref{fig:armchair} our numerical results for the band gap in an armchair-like nanoribbon ($\varphi=0$). We see that the gap oscillates with a period of 1.5 unit cells, in agreement with Eq.\ \eqref{eq:armdelta}.
	
\section{Conclusion}
\label{conclusion}
\begin{figure}[tbh]
\includegraphics[width=0.8\linewidth]{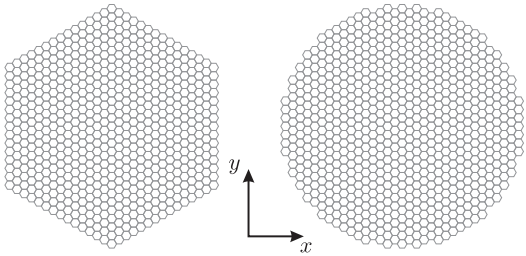}
\caption{\label{wow} 
These two graphene flakes (or quantum dots) both have the same zigzag-type boundary condition: $\Psi=\pm\tau_{z}\otimes\sigma_{z}\Psi$. The sign switches between $+$ and $-$ when the tangent to the boundary has an angle with the $x$-axis which is a multiple of $60^{\circ}$.
}
\end{figure}

	In summary, we have demonstrated that the zigzag-type boundary condition $\Psi= \pm \tau_z\otimes\sigma_z \Psi$ applies generically to a terminated honeycomb lattice. The boundary condition switches from the plus-sign to the minus-sign at the armchair orientation $\varphi=0\ (\textrm{modulo } \pi/3)$, when the boundary is parallel to $1/3$ of all the carbon-carbon bonds (see Fig.\ \ref{wow}).

	The distance $\delta$ from the edge within which the boundary condition breaks down is minimal ($=0$) at the zigzag orientation $\varphi=\pi/6\ (\textrm{modulo } \pi/3)$ and maximal at the armchair orientation. This is the length scale that governs the band gap $\Delta\approx(\hbar v/W)\exp(-W/\delta)$ in a nanoribbon of width $W$. We have tested our analytical results for $\Delta$ with the numerical solution of the tight-binding equations and find good agreement.
	
	While the lattice termination by itself can only produce zigzag or armchair-type boundary conditions, other types of boundary conditions can be reached by breaking the electron-hole symmetry of the tight-binding equations. We have considered the effect of a staggered potential at a zigzag boundary (produced for example by edge magnetization \cite{Son06}), and have calculated the corresponding boundary condition. It interpolates smoothly between the zigzag and infinite-mass boundary conditions, opening up a gap in the spectrum that depends on the strength and range of the staggered potential.

	We have calculated the dispersion relation for arbitrary boundary conditions and found that the edge states which are dispersionless at a zigzag edge acquire a dispersion for more general boundary conditions. Such propagating edge states exist, for example, near a zigzag edge with staggered potential.

	Our discovery that the zigzag boundary condition is generic explains the findings of several computer simulations \cite{Nak96,Bee07,Ryc07} in which behavior characteristic of a zigzag edge was observed at non-zigzag orientations. It also implies that the mechanism of gap opening at a zigzag edge of Ref.\ \cite{Son06} (production of a staggered potential by magnetization) applies generically to any $\varphi\neq 0$. This may explain why the band gap measurements of Ref.\ \cite{Han07} produced results that did not depend on the crystallographic orientation of the nanoribbon.

\acknowledgements

This research was supported by the Dutch Science Foundation NWO/FOM. We acknowledge helpful discussions with I. Adagideli, J. H. Bardarson, Ya. B. Bazaliy, and I. Snyman.

\appendix
\section{Derivation of the general boundary condition (\ref{MainCond})\label{CondCalc}}

We first show that the anticommutation relation (\ref{ConditionMJ}) follows from the current conservation requirement (\ref{CurrentAbsence}). The current operator in the basis of eigenvectors of $M$ has the block form
\begin{equation}
\bm{n}_B \cdot \bm{J}=
\begin{pmatrix}
{X}&{Y}\\
{Y^{\dagger}}&{Z}
\end{pmatrix},\;\;
M=\begin{pmatrix}
1&0\\
0&-1
\end{pmatrix}.
\end{equation}
The Hermitian subblock $X$ acts in the two-dimensional subspace of eigenvectors of $M$ with eigenvalue $1$. To ensure that $\langle\Psi|\bm{n}_{B}\cdot\bm{J}|\Psi\rangle=0$ for any $\Psi$ in this subspace it is necessary and sufficient that $X=0$. The identity $(\bm{n}_B \cdot \bm{J})^2=1$ is equivalent to $YY^{\dagger}=1$ and $Z=0$, hence $\{M, \bm{n}_B \cdot \bm{J}\}=0$.

We now show that the most general $4\times 4$ matrix $M$ that satisfies Eqs.\ \eqref{ConditionM1} and \eqref{ConditionMJ} has the 4-parameter form \eqref{MainCond}. Using only the Hermiticity of $M$, we have the 16-parameter representation
\begin{equation}
M=\sum_{i,j=0}^{3}(\tau_{i}\otimes\sigma_{j})c_{ij},\label{M16}
\end{equation}
with real coefficients $c_{ij}$. Anticommutation with the current operator brings this down to the 8-parameter form
\begin{equation}
M=\sum_{i=0}^{3}\tau_{i}\otimes (\bm{n}_{i}\cdot\bm{\sigma}),
\end{equation}
where the $\bm{n}_i$'s are three-dimensional vectors orthogonal to $n_B$. The absence of off-diagonal terms in $M^2$ requires that the vectors $\bm{n}_1,\,\bm{n}_2,\,\bm{n}_3$ are multiples of a unit vector $\tilde{\bm{n}}$ which is orthogonal to $\bm{n}_0$. The matrix $M$ may now be rewritten as
\begin{equation}
M=\tau_0\otimes (\bm{n}_0\cdot\bm{\sigma})+(\tilde{\bm{\nu}}\cdot\bm{\tau})\otimes(\tilde{\bm{n}}\cdot\bm{\sigma}).
\end{equation}
The equality $M^2=1$ further demands $\bm{n}_0^2+\tilde{\bm{\nu}}^2=1$, leading to the 4-parameter representation (\ref{MainCond}) after redefinition of the vectors.

\section{Derivation of the boundary modes}
\label{App:modes}
We derive the characteristic equation \eqref{eq:chareq} from the tight-binding equation \eqref{eq:TBHam} and the definitions of the boundary modes  \eqref{eq:blochdef} and \eqref{eq:lambdadef}. In the low energy limit $\varepsilon/t\ll a/|\bm T|$ we may set $\varepsilon\rightarrow 0$ in Eq.\ \eqref{eq:TBHam}, so it splits into two decoupled sets of equations for the wave function on sublattices $A$ and $B$:
\begin{subequations}
\begin{gather}
\psi_B(\bm r)+\psi_B(\bm r -\bm R_1)+\psi_B(\bm r -\bm R_2)=0,\\
\psi_A(\bm r)+\psi_A(\bm r +\bm R_1)+\psi_A(\bm r +\bm R_2)=0.
\end{gather}
\end{subequations}
Substituting $\bm R_1$ by $\bm R_2+ \bm R_3$ in these equations and using the definition \eqref{eq:lambdadef} of $\lambda$ we express $\psi(\bm r+\bm R_2)$ through $\psi(\bm r)$,
\begin{subequations}
\label{eq:lambdamode1}
\begin{align}
\psi_B(\bm r +\bm R_2)&=-(1+\lambda)^{-1}\psi_B(\bm r),\\
\psi_A(\bm r +\bm R_2)&=-(1+\lambda)\psi_A(\bm r).
\end{align}
\end{subequations}
Eqs.\ \eqref{eq:lambdadef} and\ \eqref{eq:lambdamode1} together allow to find the boundary mode with a given value of $\lambda$ on the whole lattice:
\begin{subequations}
\label{eq:lambdamode2}
\begin{align}
\psi_B(\bm r +p \bm R_2 + q \bm R_3)&=\lambda^{q}(-1-\lambda)^{-p}\psi_B(\bm r),\\
\psi_A(\bm r +p \bm R_2 + q \bm R_3)&=\lambda^{q}(-1-\lambda)^{p}\psi_A(\bm r),
\end{align}
with $p$ and $q$ arbitrary integers.
\end{subequations}
Substituting $\psi(\bm r +\bm T)$ into Eq.\ \eqref{eq:blochdef} from Eq.\ \eqref{eq:lambdamode2} and using $\bm T=(n+m)\bm R_2+n\bm R_3$ we arrive at the characteristic equation \eqref{eq:chareq}.

\begin{figure}[tbh]
\includegraphics[width=0.8\linewidth]{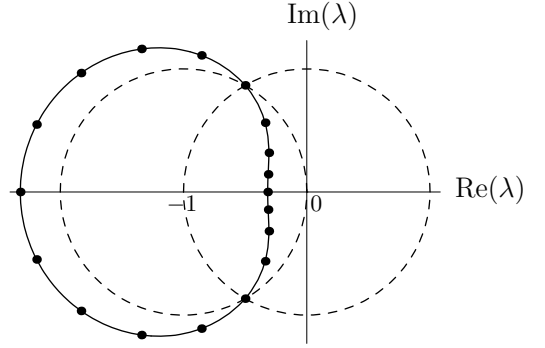}
\caption{\label{fig:lambda} Plot of the solutions of the characteristic equations (\ref{eq:absval}, \ref{eq:arg}) for $n=5$, $m=11$, and $k=0$. The dots are the roots, the solid curve is the contour described by Eq.\ \eqref{eq:absval}, and the dashed circles are unit circles with centers at $0$ and $-1$.}
\end{figure}

  We now find the roots of the Eq.\ \eqref{eq:chareq} for a given $k$. It is sufficient to analyze the equation for sublattice $A$ only since the calculation for sublattice $B$ is the same after interchanging $n$ and $m$. The analysis of Eq.\ \eqref{eq:chareqA} simplifies in polar coordinates,
\begin{gather}
|1+\lambda|^{m+n}=|\lambda|^{n}\label{eq:absval}\\
(m+n)\arg(-1-\lambda)-k-n\arg(\lambda)= 2 \pi l \label{eq:arg},
\end{gather}
with $l=0,\pm 1,\pm2\ldots$. The curve defined by Eq.\ (\ref{eq:absval}) is a contour on the complex plane around the point $\lambda=-1$ which crosses points $\lambda_\pm=-1/2\pm i\sqrt{3}/2$ (see Fig. \ref{fig:lambda}). The left-hand side of Eq.\ (\ref{eq:arg}) is a monotonic function of the position on this contour. If it increases by $2\pi \Delta l$ on the interval between two roots of the equation, then there are $\Delta l-1$ roots inside this interval. For $k=0$ both $\lambda_{-}$ and $\lambda_{+}$ are roots of the characteristic equation. So in this case the number ${\cal N}_A$ of roots lying inside the unit circle can be calculated from the increment of the left-hand side of Eq.\ (\ref{eq:arg}) between $\lambda_{-}$ and $\lambda{+}$:
\begin{equation}
{\cal N}_A=\frac{1}{2 \pi}\left[ (n+m)\frac{2 \pi}{3}+ n\frac{2 \pi}{3} \right]-1 = n - \frac{n-m}{3} - 1.
\end{equation}
Similarly, on sublattice $B$, we have (upon interchanging $n$ and $m$),
\begin{equation}
{\cal N}_B = m - \frac{m-n}{3} - 1.
\end{equation}
   
   The same method can be applied to calculate $\lambda_n$. Since there are $n-1$ roots on the contour defined by Eq.\ \eqref{eq:absval} between $\lambda_n$ and $\lambda_n^{\ast}$, the increment of the left-hand side of Eq.\ \eqref{eq:arg} between $\lambda_n^{\ast}$ and $\lambda_n$ must be equal to $2\pi(n-1)\approx 2\pi n$ (for $|\bm T|\gg a$), which immediately leads to Eq.\ \eqref{eq:lambdan} for $\lambda_n$.

\end{document}